\documentclass[structabstract]{aa}  

\usepackage{natbib}
\usepackage{amssymb,amsfonts,amstext}
\usepackage{amsmath}

\newcommand{\Ha}{${\rm H}\alpha$}
\newcommand{\Hb}{${\rm H}\beta$}

\bibpunct{(}{)}{;}{a}{}{,} 

\usepackage{graphicx}

\begin{document}
   \title{Accretion and activity on the post-common-envelope binary \object{RR~Cae}.\footnote{Based on data products from observations made with ESO Telescopes at the La Silla Paranal Observatory under program ID 076.D-0142.}}

   \institute{Universidade Federal de Sergipe, Centro de Ci\^encias Exatas e Tecnologia, Departamento de F\'isica, Cidade Universit\'aria Prof. Jos\'e Alo\'isio de Campos, Rod. Marechal Rondon s/n, Jardim Rosa Elze, S\~ao Crist\'ov\~ao, SE, Brasil. ZIP: 49.100-000 \email{tribeiro@ufs.br}\label{inst1}
\and
SOAR Telescope \label{instUFS}
         \and
         Universidade Federal de Santa Catarina \email{bap@astro.ufsc.br}\label{inst2}
         \and
             Department of Terrestrial Magnetism \email{skafka@ciw.edu}\label{inst3}
          \and
             D\'epartement de physique - Universit\'e de Montr\'eal C.P. 6128, Succ. Centre-Ville, Montr\'eal, Quebec H3C 3J7, Canada; \email{dufourpa@astro.umontreal.ca} \email{fontaine@astro.umontreal.ca}\label{inst4}
           \and
	    Homer L. Dodge Department of Physics and Astronomy University of Oklahoma, 440 W. Brooks St., Norman, OK, 73019, USA \email{alexg@nhn.ou.edu}\label{inst5}
             }

   \author{Ribeiro, T.
          \inst{\ref{inst1},\ref{instUFS}}
          \and
	Baptista, R.
	\inst{\ref{inst2}}
	\and
	Kafka, S.
	\inst{\ref{inst3}}
	\and
	Dufour, P.
	\inst{\ref{inst4}}
	\and
	Gianninas, A.
	\inst{\ref{inst4},\ref{inst5}}
	\and
	Fontaine, G.
	\inst{\ref{inst4}}
          }

   \date{Received ??? ??, ????; accepted ??? ??, ????}

  \abstract
   {
   Current scenarios for the evolution of interacting close binaries -- such as cataclysmic variables (CVs) -- rely mainly on our understanding of low-mass star angular momentum loss (AML) mechanisms. The coupling of stellar wind with its magnetic field, i.e., magnetic braking, is the most promising mechanism believed to drive AML in these stars. There are basically two properties thought to drive magnetic braking: the stellar magnetic field and the stellar wind. Understanding the mechanisms that drive AML therefore requires a  comprehensive understanding of these two properties as well.
   }
   {
   \object{RR~Cae} is a well-known nearby ($d = 20 \rm pc$) eclipsing DA+M binary with an orbital period of $P = 7.29\rm h$. The system harbors a metal-rich cool DA white dwarf (WD) and a highly active M-dwarf locked in synchronous rotation. The metallicity of the WD suggests that wind accretion is taking place, which provides a good opportunity to obtain the mass-loss rate of the M-dwarf component. We aim to reach a better understanding of the AML mechanisms in close binaries by characterizing the relevant properties of the M-dwarf component of this system.
   }
   {
   We analyzed multi-epoch time-resolved high-resolution spectra of \object{RR~Cae} in search for traces of magnetic activity and accretion. We selected a number of well-known chromospheric activity indicators and studied their phase-dependence and long-term behavior. Indirect-imaging tomographic techniques were also applied to provide the surface brightness distribution of the magnetically active M-dwarf. The blue part of the spectrum was modeled using a state-of-the-art atmosphere model to constrain the WD properties  and its metal enrichment. The latter was used to improve the determination of the mass-accretion rate from the M-dwarf wind.
   }
   {
   Doppler imaging of the M-dwarf component of \object{RR~Cae} reveals a polar feature similar to those observed in fast-rotating solar-type stars. Analysis of tomographic reconstruction of the \Ha~emission line reveals two components, one traces the motion of the M dwarf and is generated by chromospheric activity, while the other clearly follows the motion of the WD. The presence of metals in the WD spectrum suggests that this component arises from accretion of the M-dwarf wind. A model fit to the WD spectrum provides $\rm T_{eff} = (7260\pm250)K$ and $\log g = (7.8 \pm 0.1) {\rm dex}$ with a metallicity of $<\log[X/X_\odot]> = (-2.8 \pm 0.1) {\rm dex}$. This maps into a mass-accretion rate of $\rm \dot{M}_{acc} = (7\pm2) \times 10^{-16} M_\odot \cdot yr^{-1}$ onto the surface of the WD.
   }
   {}


   \maketitle
%

\section{Introduction}
\label{sec:intro}

Detached white dwarf (WD)-main sequence (MS) binaries are probably the byproduct of an intermediate evolutionary stage of binary stars \citep{Davis:2010p2122}. These systems emerge from a common-envelope (CE) phase and further evolve into a semi-detached configuration, e.g. cataclysmic variables (CVs), where the MS component transfers matter to the WD via Roche-lobe overflow \citep{1995CAS....28.....W}.

The evolution from the CE to CV phase of these binaries is governed by angular momentum loss (AML) mechanisms: either magnetic braking from the MS component or gravitational radiation. In fact, the AML mechanisms that drive the system to the CV phase probably also drive the subsequent stages of evolution.  

While gravitational radiation depends mainly on well-defined parameters \citep{Schreiber:2003p128}, magnetic-braking models are mostly based on parametrized or empirical descriptions \citep{Barnes:2010p3060} that may harbor considerable uncertainty. Clearly, understanding the properties of the magnetically active MS component in WD+MS binaries is crucial for a quantitative description of the evolution of close binaries \citep{Davis:2010p2122, Davis:2008p2123,2011ApJS..194...28K}.

The detection of metal lines in the spectra of WD-binaries in a wide range of orbital separations \citep[$a\sim 10^{-2} - 10^{-3} \rm AU$ ]{2011A&A...532A.129T,Debes:2006p3314,2003ApJ...596..477Z} suggests that low-level accretion takes place much earlier than the start of the CV phase. It is very likely that material from the wind of the MS component is accreted onto the surface of the WD, which results in the observed metal enrichment. \cite{Debes:2006p3314} was able to constrain the mass-loss rate of low-mass star companions by studying the metal enrichment of WDs in close binaries. Such studies not only improve our understanding of the dynamics of low-mass stellar winds but also place constrains on their braking timescales, thus enabling a better understanding the evolution of close binaries.

The source \object{RR~Cae} (\object{WD~0419-487}) is a nearby ($d = 20 \rm pc$, \citealt{2009AJ....137.4547S}), detached, eclipsing DA+M binary with an orbital period of $\rm P_{orb} = 7.29\rm h$. The M-dwarf component is tidally locked to the system, displaying strong chromospheric emission and enhanced magnetic activity \citep{Bruch:1998p4261, Bruch:1999p3205} as well as coronal x-ray emission \citep{2010AJ....140.1433B}. Photospheric metal features associated to the WD were identified in the spectrum of the system by \cite{2003ApJ...596..477Z} which, together with the shape of the x-ray spectrum of the system \citep{2010AJ....140.1433B}, supports the idea that low-level accretion is taking place, probably from the M-star wind. \citet{Maxted:2007p4248} provided an extensive analysis of spectroscopic and photometric data, improving the previous determination of the system  parameters by \cite{Bruch:1998p4261} and \cite{Bruch:1999p3205}.  

Here we present a detailed analysis of archival multi-epoch time-resolved spectroscopic data of \object{RR~Cae}, aimed at a better understanding of the magnetic properties of its M component and the wind-accretion scenario. In Sect.~\ref{sec:obs} we describe the data set, review the reduction procedures and epoch sorting. The analysis of the data is presented in Sect.~\ref{sec:anl}. We start by measuring the radial velocity of the binary components from different emission/absorption features and by analyzing magnetic activity indicators of the M component. Doppler-imaging of the M star is also presented, adding \object{RR~Cae} to the number of fast-rotating low-mass stars with surface images reconstructions. We measure the WD metal enrichment due to accretion, with the aid of a state-of-the-art model atmosphere. In Sect.~\ref{sec:disc} we discuss our results in view of similar analyses performed on different objects and previous analyses of the \object{RR~Cae} system. In Sect.~\ref{sec:end} we present our final remarks and a summary of the results.

\section{Observations and data reduction}
\label{sec:obs}

\begin{table}
\caption{\label{tab:obslog} Log of observations of RR~Cae with UVES/VLT. Labels in boldface indicate epochs of enhanced stellar activity (see text for discussion). }
\centering
\begin{tabular}{lccc}
\hline\hline
JD (start)	                &	Cycle		&	Phase	 &	Label		\\
\hline
$2\,453\,658.25$		&	$7032  $	&$[0.19 : 0.30]$ & {\bf EP01} \\
$2\,453\,659.16$		&	$7035  $	&$[0.19 : 0.30]$ & {\bf EP02} \\
$2\,453\,660.25$		&	$7038  $	&$[0.80 : 0.92]$ & EP03 \\
$2\,453\,661.23$		&	$7042  $	&$[0.01 : 0.12]$ & EP04 \\
$2\,453\,662.26$		&	$7045  $	&$[0.40 : 0.50]$ & EP05 \\
$2\,453\,663.18$		&	$7048  $	&$[0.44 : 0.54]$ & EP06 \\
$2\,453\,665.16$		&	$7054+1$	&$[0.95 : 1.05]$ & EP07 \\
$2\,453\,666.20$		&	$7058  $	&$[0.38 : 0.48]$ & {\bf EP08} \\
$2\,453\,667.22$		&	$7061  $	&$[0.73 : 0.83]$ & EP09 \\
$2\,453\,671.21$		&	$7074  $	&$[0.87 : 0.98]$ & {\bf EP10} \\
$2\,453\,676.30$		&	$7091  $	&$[0.65 : 0.75]$ & EP11 \\
$2\,453\,679.07$		&	$7100  $	&$[0.76 : 0.87]$ & {\bf EP12} \\
$2\,453\,680.16$		&	$7104  $	&$[0.34 : 0.45]$ & {\bf EP13} \\
$2\,453\,686.10$		&	$7123+1$	&$[0.91 : 1.02]$ & EP14 \\
$2\,453\,687.09$		&	$7127  $	&$[0.16 : 0.26]$ & EP15 \\
$2\,453\,690.23$		&	$7137  $	&$[0.50 : 0.61]$ & EP16 \\
\hline
\end{tabular}

\end{table}

Observing runs were performed on successive nights from 2005 October 14 to 23, and at different time intervals thereafter. We labeled the observing runs as EP01 to EP16 in chronological order, highlighting those with evidence of increased magnetic activity in the M-dwarf component (see analysis below). For each visit, a total of eight spectra with exposure times of $340s$ was acquired. 

Data reduction was performed with the ESO/Gasgano UVES pipeline (version 3.0.1) and included bias- and flat-field corrections and wavelength calibration with a ThAr lamp. Flux calibration was obtained from a master sensitivity function derived from observations of standard stars. { The resulting flux calibration provided by this step is not as reliable as that provided by contemporaneous observations of standard stars. Nevertheless, this should not affect our analysis as it does not strongly rely on an accurate flux calibration. It is also interesting to note that the flux calibration with master sensitivity function leads to better results for the redder part of the spectrum while providing quite poor results in the blueward regions. }

The data were phase-folded according to photometric the linear ephemeris
\begin{equation}
\label{eq:efem}
{\rm MJD} = 2\,451\,522.548\,5670(19) + 0.303\,703\,6366(47) \cdot {\rm E}
\end{equation}
provided by \cite{Maxted:2007p4248}, which assumes phase zero as mid-eclipse of the WD. We list the phase coverage of each observing run in Table~\ref{tab:obslog}. 

\section{Data analysis}
\label{sec:anl}

\begin{figure*}
\centering
  \includegraphics[width=0.45\textwidth]{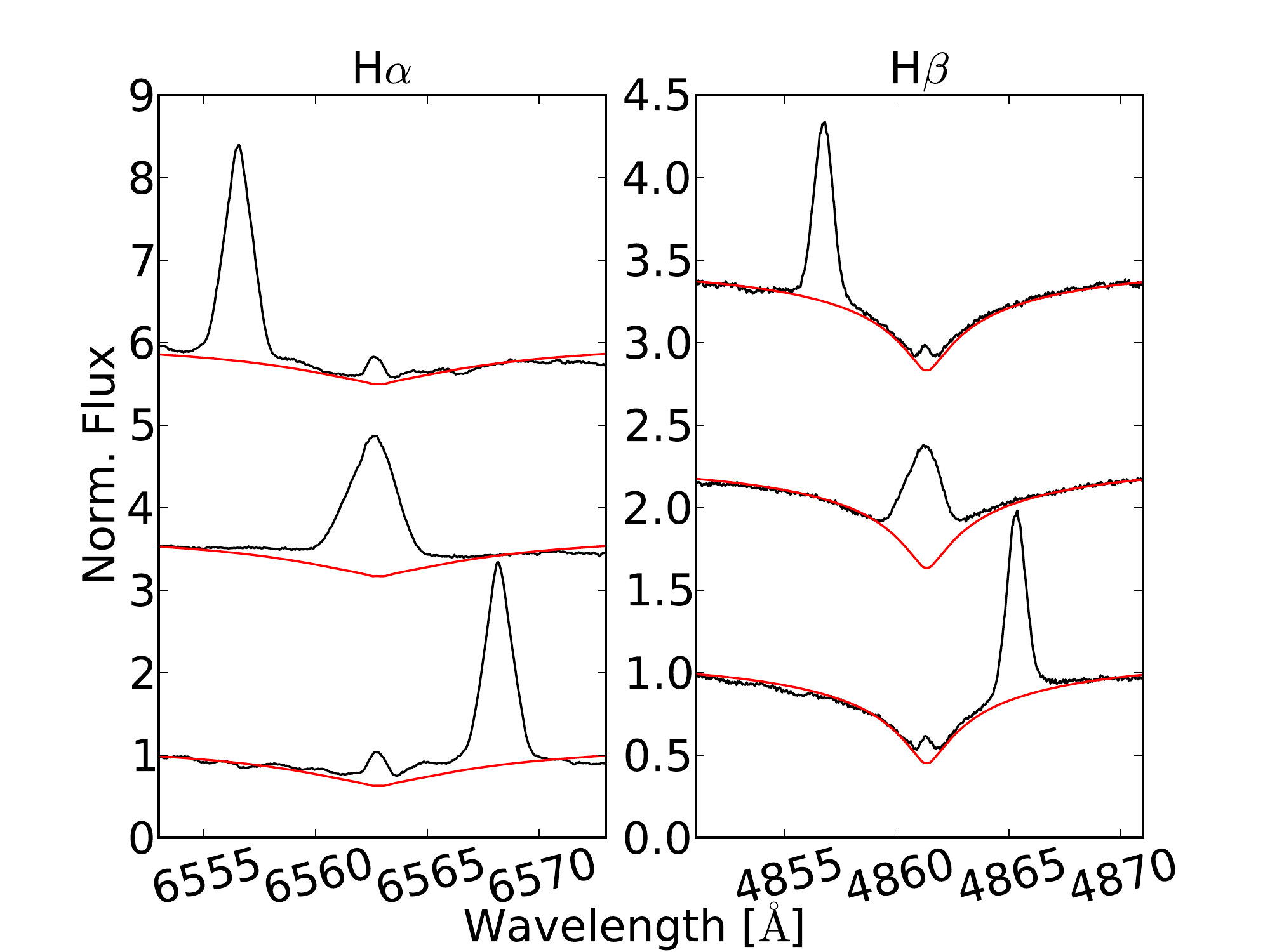}
  \includegraphics[width=0.45\textwidth]{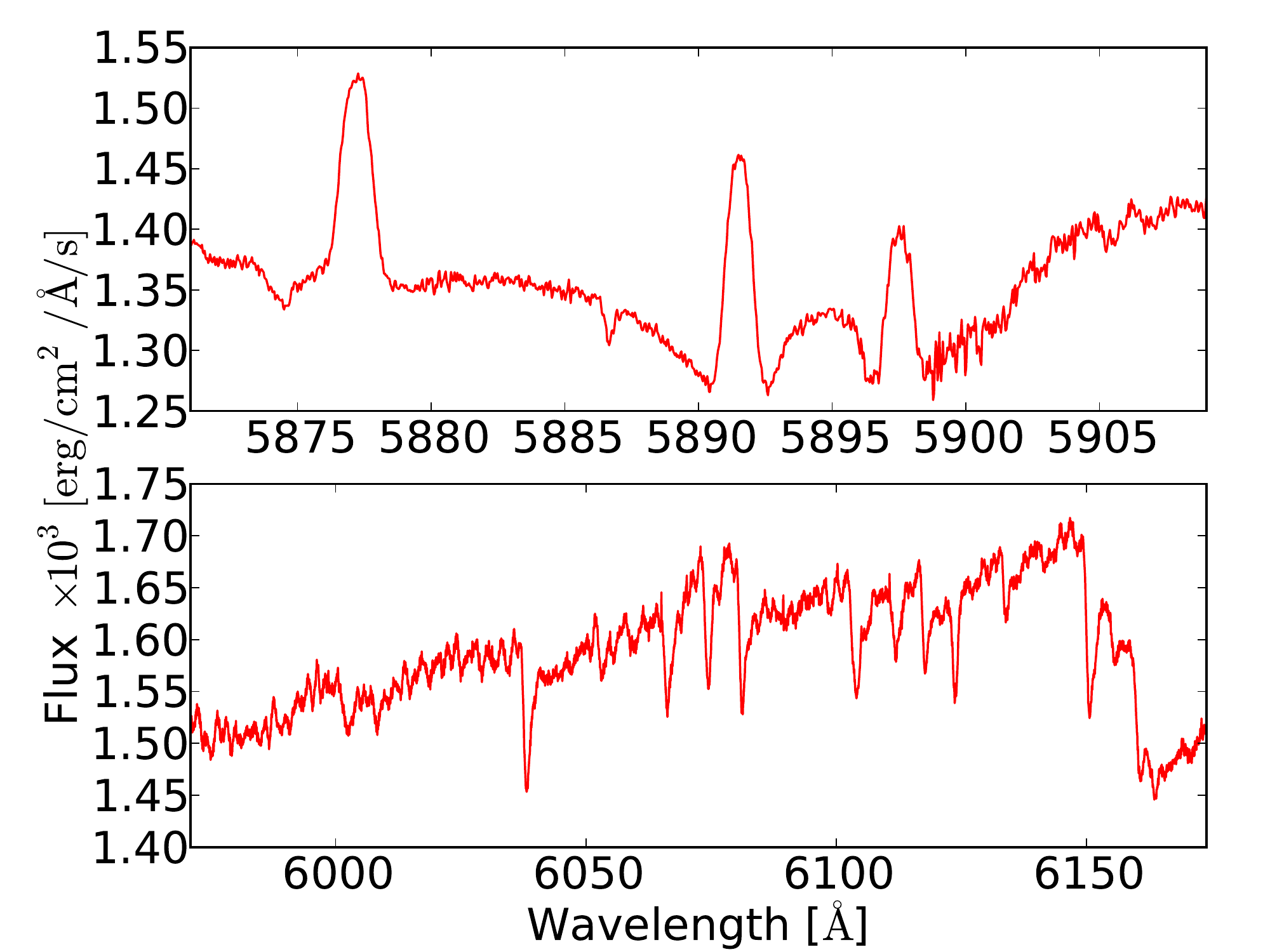}
  
  \includegraphics[width=0.85\textwidth]{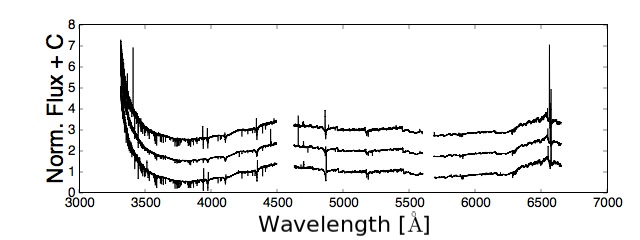}
    \caption{Sample spectroscopic regions of the optical spectra of \object{RR~Cae}. The {\it left} two panels show a combinations of all spectra around quadratures (phases [0.2:0.3] and [0.7:0.8], top and bottom, respectively) and superior conjunction (phases [0.45:0.55], middle spectra) around \Ha~and \Hb. The spectra where rectified by the radial velocity of the WD so that the center of the broad Balmer absorption matches the rest wavelength of the line. { A red line represents the WD model spectrum scaled to the appropriate relative contribution of the star and is vertically shifted to match the continuum level (see text for discussion). The spectra at each phase are also shifted in the y direction for better visualization.} At the {\it right-hand side} we show a combination of spectra of the M-dwarf component alone (taken during the eclipse of the WD), around the \ion{Na}{I} doublet ($\rm D1 = 5896\AA$ and $\rm D2 = 5890\AA$) and \ion{He}{I} ($\rm 5876\AA$) region, and in a region with abundant photospheric metal lines from the M dwarf, at the top and bottom panels, respectively. { An average of all spectra around quadratures and superior conjunction is shown at the {bottom panel}. Each spectrum is divided by the mean of the spectrum between $4575 \rm \AA$ and $5540 \rm \AA$ and is vertically shifted for better visualization.}}
   \label{fig:sp2_H_avg}
   \label{fig:sp2_redu_avg}
\end{figure*}

{ To highlight the main features from the spectra of \object{RR~Cae} we selected and averaged all spectra around quadratures and superior conjunction (phases 0.25, 0.75 and 0.5, respectively). The result is shown in the bottom panel of Fig.~\ref{fig:sp2_H_avg}. As we mentioned earlier, the spectra were flux calibrated using a master sensitivity function, which may be quite unreliable in the blue. The photometric calibration probably has only a marginal impact (if any) on our analysis. In any case, the majority of the interesting spectroscopic features lies in regions with good photometric calibration.}

{ We point out a number of prominent  metal absorption lines in the blueward region of the spectra that arise from the WD, as revealed by a radial velocity analysis. It is also possible to detect multiple components to the Balmer emission lines, specially \Ha~and \Hb, as shown in the upper left panels of Fig.~\ref{fig:sp2_H_avg}. Absorption and emission features from the highly active M dwarf are also present in the red part of the spectrum and can be seen in the inset at the upper right panels of Fig.~\ref{fig:sp2_H_avg}.}

\subsection{Radial velocities}
\label{ssec:rv}

Obtaining radial velocity (RV) measurements for binary stars is paramount for determining accurate system parameters as well for determining the origin of the different components of the spectrum. We proceeded by measuring the RVs of different spectral features to all spectra, which where then phase-folded according to the linear ephemeris of Eq.~\ref{eq:efem} and used in a fit to a sinusoid, 
\begin{equation}
\label{eq:rv}
v_r(\phi) = K \sin [ 2 \pi  (\phi + \delta\phi) ] + \gamma,
\end{equation}
to yield the semi-amplitude of the RV ($K$), the systemic velocity ($\gamma$), and a phase displacement ($\delta\phi$). Since \object{RR~Cae} is an eclipsing system, the ephemeris (orbital period and phase of conjunction, $\phi = 0$) is well stablished, justifying the adoption of this simple form. We kept the phase displacement to account for possible small-scale variations in the ephemeris, as reported by \cite{2010MNRAS.407.2362P}.

We started by fitting the main features of the Balmer lines, namely the emission line contribution of the active MS star and (when required) the broad absorption from the WD using Gaussian profiles. For instance, for the  \Ha~emission line from the MS component, a single Gaussian  provided a good fit to the data, while for \Hb~and higher order H Balmer lines a second Gaussian was required to fit the broad absorption feature from the WD. 

\begin{figure}
\centering
\includegraphics[width=0.5\textwidth]{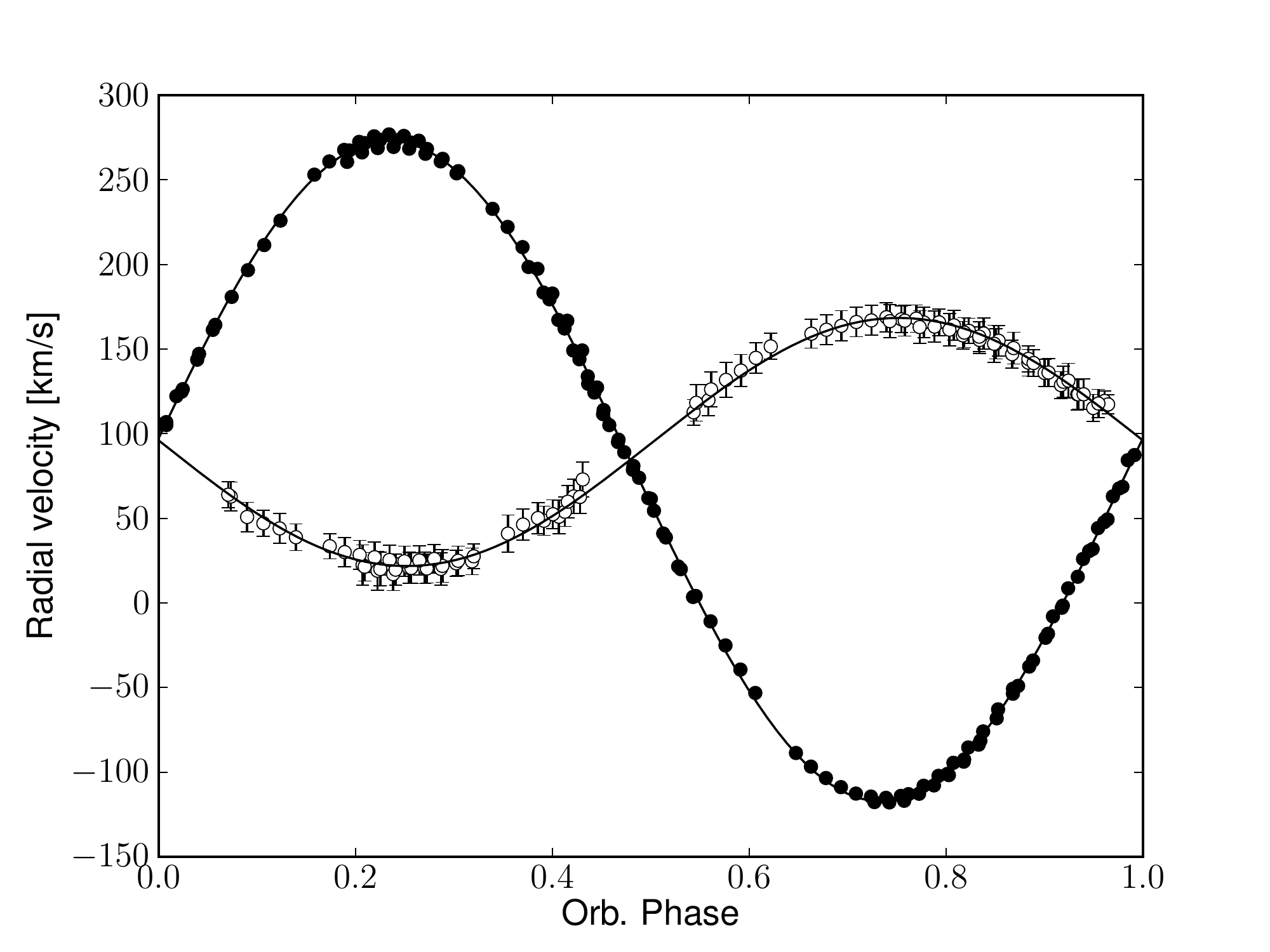}
   \caption{Radial velocity of the \Ha~emission line of \object{RR~Cae}. Filled symbols represent the measurement of the \Ha~prominent from the M dwarf component. Open symbols with error bars represents the measurement from the extra emission component. The uncertainty in the data points of the M dwarf radial velocity is of the order of the size of the symbols and, therefore, are not shown in the fit. Solid lines represent least-square sinusoidal fits to the two emission components  data.}
\label{fig:rv}
\end{figure}

To fit the additional emission line component observed in the spectra of \object{RR~Cae} (Fig.~\ref{fig:sp2_H_avg}) we performed an interactive procedure using the FITPROFS task from the ONEDSPEC package of IRAF\footnote{IRAF is distributed by the National  Optical Astronomy Observatories, which are operated by the Association of  Universities for Research in Astronomy, Inc., under cooperative agreement with the National Science Foundation.}, assuming two Gaussian profiles. In this case, a selection of spectra around quadratures (where the emission lines are not blended) were used to measure the RV and other quantities (see Sect.~\ref{ssec:act} below) of these components. Here we concentrate our efforts on the \Ha~and \Hb~lines since they are the only ones with a sufficiently strong signal to allow us to employ this deblending procedure. In Fig.~\ref{fig:rv} we show the results for the \Ha~line with filled and open symbols representing the M dwarf and the additional component, respectively.

The deblended components were fitted with the sinusoidal function of Eq.~\ref{eq:rv}. The resulting $K$ and $\gamma$ velocities for the \Ha, \Hb, and \ion{He}{I}($5876$\AA) lines are shown in Table~\ref{tab:rv}. { We employed a Markov chain Monte Carlo procedure \citep{Patil:2010tx} for fitting the data, which allowed us to reliably estimate the uncertainty in the fit -- the numbers within parentheses in Table~\ref{tab:rv}.} The continuous line in Fig.~\ref{fig:rv} represents these fits for the multicomponent \Ha~line.

\begin{table}[h]
\caption{\label{tab:rv} Radial velocity solutions for the identified spectral features in the spectra of RR~Cae. The values within parenthesis are the uncertainty in the last digit.}
\centering
\begin{tabular}{lrrc}
\hline\hline
Feature		 		& 	$K [km/s]$		 &	$\gamma [km/s]$	&	$\delta\phi$\\
\hline
${\rm H}\alpha({\rm MS})$\tablefootmark{a} 	& $195.1(3)$ &	$77.9(4)$	&	$-0.0160(3)$ \\
${\rm H}\alpha({\rm EX})$\tablefootmark{b} 	& $73.3 (4)$ &	$95.1(3)$	&	$-0.018(1)$ \\
${\rm H}\beta({\rm MS})$\tablefootmark{a} 	& $195.2(3)$ &	$80.5(4)$	&	$-0.0161(2)$ \\
${\rm H}\beta({\rm EX})$\tablefootmark{b} 	 	& $75(1)$ &	$91.0(7)$	&	$-0.011(2)$ \\
${\rm H}\beta({\rm WD})$\tablefootmark{c}		& $74(2)$ &	$98(1)$	&	$-0.014(4)$ \\
\ion{He}{I}($5876\rm\AA$)\tablefootmark{a} 	& $193.1(5)$ &	$60.3(3)$	&	$-0.0171(4)$ \\
$[4700:5597]\rm\AA$\tablefootmark{d}		& $ 194.8(2)$ &$77.5(7)$	&	$-0.0151(2)$ \\
LSD Profile							& $ 193.9(2)$ &$82.2(1)$	&	$-0.0158(1)$\\	
\hline
\end{tabular}
\tablefoot{ \\
\tablefoottext{a}{From the MS star.}
\tablefoottext{b}{Extra component.}
\tablefoottext{c}{Broad absorption from the WD.}
\tablefoottext{d}{Cross-correlations of M-dwarf spectra with the M4V template from \citet{Cincunegui:2004p4262}.}
}

\end{table}

In addition, we also obtained RV measurements of the M-dwarf component by cross-correlating selected regions of its spectrum with that of the template star and fitting Eq.~\ref{eq:rv} to the results to estimate $K$, $\gamma$, and $\delta\phi$. The cross-correlation procedure requires a template spectrum with  spectral type similar to that of the target. For that purpose we selected a spectrum of an M4 dwarf \citep{Maxted:2007p4248} from \citet{Cincunegui:2004p4262}. At the same time, it is important to select an appropriate spectral region containing strong spectral features to provide a reliable solution. The results of this procedure are shown in Tab.~\ref{tab:rv} together with the selected spectral regions. 

We performed a cross-check in the determination of  the RV measurements by replacing the \citet{Cincunegui:2004p4262} template star by one of the spectrum of \object{RR~Cae} during eclipse (similar to the procedure used by \citealt{2011A&A...532A.129T}). The RV semi-amplitudes resulting from this procedure are similar to that obtained by means of a template spectrum (within the $1-\sigma$ limits). Because using the spectrum of the star itself to measure the RVs does not yield a measurement of the $\gamma$-velocity or of an orbital phase displacement, we relied on the results from the cross correlation with the template procedure.

In general, the RV solutions (see Tab.~\ref{tab:rv}) of the different spectral features for the MS star agree well, both from the chromospheric emission lines and from the photospheric absorption/molecular bands. Our results are also consistent with those of \cite{Maxted:2007p4248},  \cite{Bruch:1999p3205} and \cite{Bruch:1998p4261}. Combining the RV solution for the two components of the system we obtain
\[
q = \frac{\rm M_{MS}}{\rm M_{WD}} = \frac{\rm K_{WD}}{\rm K_{MS}} = 0.376 \pm 0.005
\]
and
\[
v_{gr} = \gamma_{\rm WD} - \gamma_{\rm MS} = 20 \pm 1 ~{\rm km~\cdot~s^{-1}}
\]
for the mass ratio ($q$) and the gravitational redshift of the WD ($v_{gr}$), respectively. { Here we adopted the $\gamma$ velocities from the $\rm H\alpha(MS)$ and $\rm H\beta(WD)$ in measuring $v_{gr}$. Although $\rm H\beta(WD)$ displays a larger uncertainty -- because the broad absorption feature from the WD spectra is less prominent than these emission lines -- this component best maps the surface of the WD. By using the $\gamma$ velocities provided by the additional component to the spectra, we obtained lower values for the gravitational redshift ($17.2 \pm 0.5 ~\rm km/s$). This suggests that the emission originates at some height above the surface of the WD, which is consistent with the scenario of wind accretion we discuss in Sect.~\ref{ssec:acc}. }

We also point out the good agreement between the RV measured from the chromospheric emission line of spectra taken at different epochs. Nevertheless, the standard deviation in the phase range $0.15 - 0.30$ is clearly higher than the uncertainty in the measurements (which is on the order of the plotted points, see Fig.~\ref{fig:rv}). This is possibly a result of the magnetic activity of the MS components which caused jitter in the center of light of the star. We discuss this in more detail below.

\subsection{Activity}
\label{ssec:act}

\begin{figure*}
\centering
\includegraphics[width=0.85\textwidth]{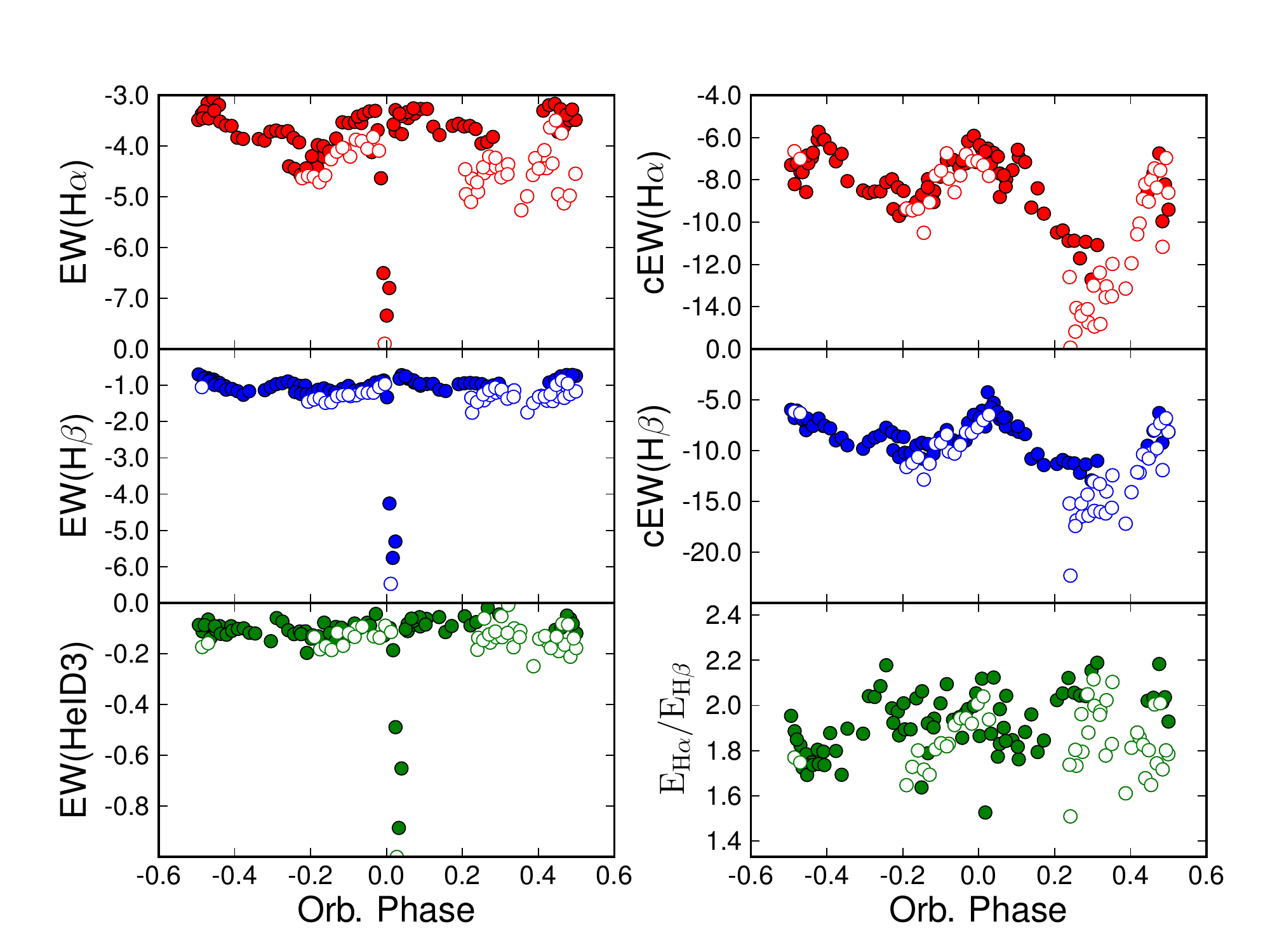}
\includegraphics[width=0.85\textwidth]{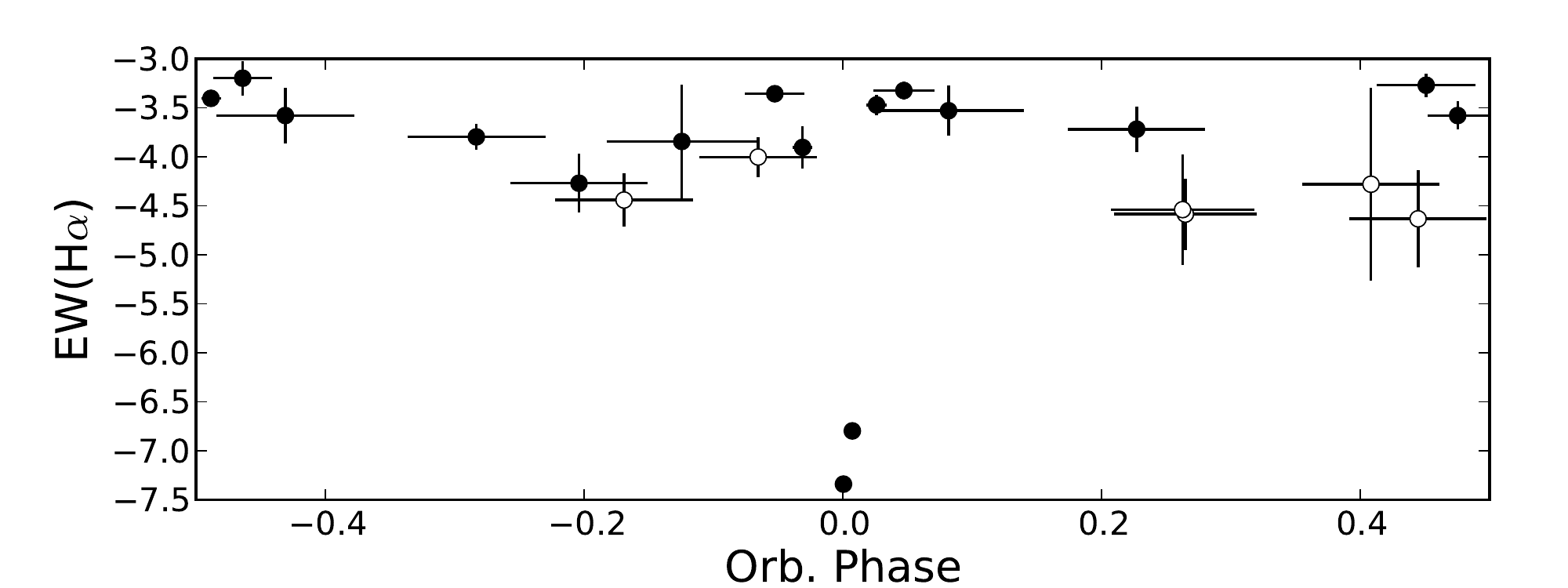}
  \caption{Phase-folded equivalent width of selected chromospheric emission lines and $\rm E_{H\alpha}/E_{H\beta}$ ratio from the M-dwarf component of \object{RR~Cae}. Here we adopted the convention that emission lines have negative EWs, so the lower the value, the stronger the line with respect to the continuum. {\bf Upper two-column plots:} Full time-resolution data. In the left panels we show the EWs of the \Ha, \Hb,  and \ion{He}{I}D3 emission lines. Open symbols represent epochs with overlapping phases where the average value of the EW of the \Ha~line differs by more than the mutual standard deviation. In the left panels we show the EW measured directly from the spectrum. In the right, the middle and top panels shows the same measurements corrected for the contribution of the WD. In the lower right panel we show the corresponding $\rm E_{H\alpha}/E_{H\beta}$ ratio. See text for discussion. {\bf Bottom single plot:} The mean EW for each cycle of the \Ha~line. The error bars represent the range (highest and lowest values) of each measurement.}
\label{fig:EW}
\end{figure*}

We estimated the magnetic activity of the MS star from the equivalent widths (EW) of well-known chromospheric activity indicators such as the hydrogen Balmer series emission lines (specially $\rm H\alpha$) as well as \ion{He}{I}\,D3, the NaD doublet ($\lambda = 5889 \rm \AA$ and $5895 \rm \AA$) and the \ion{Ca}{II} H and K ($3968\rm \AA$ and $3933 \rm \AA$) lines. The EW was measured together with the fit to the RV of each line profile, as discussed in Section~\ref{ssec:rv}. The results for the stronger emission lines (\Ha, \Hb~and \ion{He}{I}\,D3) are shown in the left-hand panel of Fig.~\ref{fig:EW}.

During the eclipse of the WD (around phase $0$) we observe an increase in the EW of all emission lines. In the out-of-eclipse spectra the WD is responsible for most of the flux, specially in the blueward regions, filling up the continuum and decreasing the EW. To avoid the contribution of the WD, we measured the EW of the lines during the eclipse. The results are shown in Table~\ref{tab:ew}. It is possible to see that the \Hb~line is stronger than \Ha, with a step decrease in EW for increasing Balmer series lines.  We also measured the emission core of the \ion{Ca}{II} H and K lines, which have similar strengths and are stronger than the Balmer emission lines.

\begin{table}[h]
\caption{\label{tab:ew} Equivalent width of emission lines from the MS component of RR~Cae measured in spectra taken during the eclipse of the WD.}
\centering
\begin{tabular}{lr|lr}
\hline\hline
Feature		 		& 	EW		& Feature	 		& 	EW		\\
\hline
${\rm H}\alpha$ 		& 	$-7.04$  	& ${\rm H}\epsilon$		& 	$-3.38$  	\\
${\rm H}\beta$ 			& 	$-8.94$  	& ${\rm H}\zeta$		& 	$-3.48$  	\\
${\rm H}\gamma$ 		& 	$-7.18$  	& \ion{Ca}{II}H			& 	$-10.08$ 	\\
${\rm H}\delta$ 		& 	$-4.87$  	& \ion{Ca}{II}K			& 	$-12.03$ 	\\
\hline
\end{tabular}

\end{table}

{To provide a better view of the orbital variation of the EW of the M-dwarf lines, we determined and extracted the contribution of the WD. Our procedure first corrected the data from the RV of the WD, and then weighted the WD contribution using a template spectrum. For this procedure we used the region around the \Hb~emission line (see Fig.~\ref{fig:sp2_H_avg}), which provides the highest ratio of the WD to M-dwarf contribution. The radial velocity of the different features where determined in Sect.~\ref{ssec:rv}, and we used the values for the $\rm H\beta(WD)$ component in Table~\ref{tab:rv}. For the template we used the model spectra of the WD obtained in Sect.~\ref{ssec:acc}. The average contribution of the WD to the \Hb~spectral region, obtained by this procedure, is $\rm <F_{WD}(H\beta)/F_{tot}> \sim 76\%$ with a standard deviation of $\sim 8\%$. The resulting EWs of the \Ha~and \Hb~lines, after removing  the WD contribution, are shown in the right-hand panels of Fig.~\ref{fig:EW}. This procedure emphasizes that the emission lines are much stronger outside the eclipse, showing a clear orbital modulation. In addition, there is no discontinuity in the EW behavior during the eclipse, indicating a good WD extraction procedure.}

It is not straightforward to characterize the orbital modulation of the EW of the emission lines. Because the data at hand comprise a collection of observation taken at different epochs, it is not easy to discern between a scenario where the orbital modulation is the result of a fixed configuration seen at different orbital phases from the alternative, where they are the consequence of a true cycle-to-cycle reconfiguration of the stellar magnetic field.

{ In an attempt to characterize the cycle-to-cycle variation of the EW, we show the mean value of the EW for each cycle in the lower panel of Fig.~\ref{fig:EW}. Data taken during the eclipse of the WD were measured separately to prevent the EW level of the out-of-eclipse measurements. The error bars in this plot represent the range of values for the duration and EW of the cycle. Open symbols depict cycles with overlapping phases where the difference in the average EW of the \Ha~line is larger than the standard deviation of the closest cycle. Here we used the actual HJD time of the observations and not the orbital phase, since there may be data points that are closer in phase but farther in time. The observed changes in the EW are reminiscent of those of small-scale stellar activity, possibly due to flaring events. This is particularly clear for the \Ha~emission line between phases $[0.15:0.5]$ where there are observations from seven different cycles\footnote{Between these phases there are observations from cycles; $7033,  7036,  7046,  7049,  7059,  7105,  7128$} (see Tab.~\ref{tab:obslog}). 
}

Furthermore, although the variations in \Hb~and \ion{He}{I}\,D3 are less pronounced than those in \Ha~ (see Fig.~\ref{fig:EW}), they are correlated. The EW measured for epochs EP01, EP02, EP08, EP10, EP12, and EP13 (open symbols in Fig.~\ref{fig:EW}) are lower than those from the remaining runs at the same phases. Even though the differences in the average EW are lower than the mutual standard deviations, they do differ at the $1-\sigma$ level. This behavior -- a direct correlation between \Ha~and \ion{He}{I} -- is commonly observed in stellar activity events for very active stars (see p.ex. \citealt{2011A&A...534A..30G}, and references therein). 

A more quantitative analysis of the overall changes in stellar activity may be obtained using a procedure suggested by \citet{1992AJ....104.1942H}: one subtracts the continuum-normalized spectrum of an active star from that of an inactive stellar "twin", and measures the excess fluxes $F_{\rm H\alpha}$ and $F_{\rm H\beta}$, in \Ha~ and \Hb, respectively. In our case, we first extracted the contribution of the WD and then normalized the continuum of the spectra. In search for an inactive template star we noted that for spectral types later than M3V, there is practically no absorption in the \Ha~and \Hb~ line cores (see \citealt{West:2011p4098} for instance). Therefore, instead of using a template star, we just subtracted a constant value ($= 1.0$) of the spectra. The ratio is then obtained by the equation
\begin{equation}
\label{eq:Ha_Hb}
\frac{\rm E_{H\alpha}}{\rm E_{H\beta}} = \left[ \frac{\rm F_{H\alpha}}{\rm F_{H\beta}} \right] \left[\frac{\rm F_{H\alpha,0}}{\rm F_{H\beta,0}} \right] \left[ 2.512^{\rm (B - R)} \right],
\end{equation}
where ${\rm F_{H\alpha,0}}$ and ${\rm F_{H\beta,0}}$ are the absolute flux densities, accounting for the energy ratio between the \Ha~and \Hb~photons, respectively, and ${\rm B}$ and ${\rm R}$ are the magnitude of the star in these passbands and accounts for a color term where these lines lies \citep{1992AJ....104.1942H}. The ratio of absolute flux densities is calculated to be 0.2444 for  \Ha~and \Hb. With the lack of appropriate measurements of B and R magnitudes for \object{RR~Cae} during the eclipse (without the contribution of the WD), we adopted typical values for M4V dwarfs, i.e.,  $(\rm B - R) = 2.3$. The results for the $\rm E_{H\alpha} / E_{H\beta}$ ratio are shown in the lower right panel of Fig.~\ref{fig:EW}.

The behavior of the $\rm E_{H\alpha} / E_{H\beta}$ ratio shows no apparent orbital modulation. Even cycles covering the same orbital period that display a considerable deviation in the EW of the \Ha~line present consistent values for the $\rm E_{H\alpha} / E_{H\beta}$. According to \cite{1992AJ....104.1942H} interpretation, high values (${\rm E_{H\alpha} / E_{H\beta}} > 3$) are only possible for prominence-like features. For \object{RR~Cae} we note that $\rm E_{H\alpha} / E_{H\beta} < 2.5$ at all epochs, suggesting that the M dwarf is dominated by plage-like magnetic structures. 

It is also important to note that irradiation of the M star by the WD may introduce an orbital modulation in the EW of the lines. Nevertheless, considering the temperature of the WD ($\rm T_{eff} \sim 8000 K$, see below) and the distance between the surface of the two components ($\sim a-{\rm R_2} \sim 0.8 {\rm R_\odot}$), the net heating effect would cause a minor temperature difference of $\lesssim 1.0 K$. This follows from assuming blackbody radiation for both the WD and M star (see \citealt{2009ebs..book.....K} for a more detailed consideration).

To provide a better view of the orbital phase evolution of the chromospheric lines, we applied Doppler tomography techniques to the \Ha~emission line of the system. We chose the \Ha~line alone since it is less affected by the broad absorption feature of the WD than the other Balmer lines.

\begin{figure}
\centering
\includegraphics[width=0.5\textwidth]{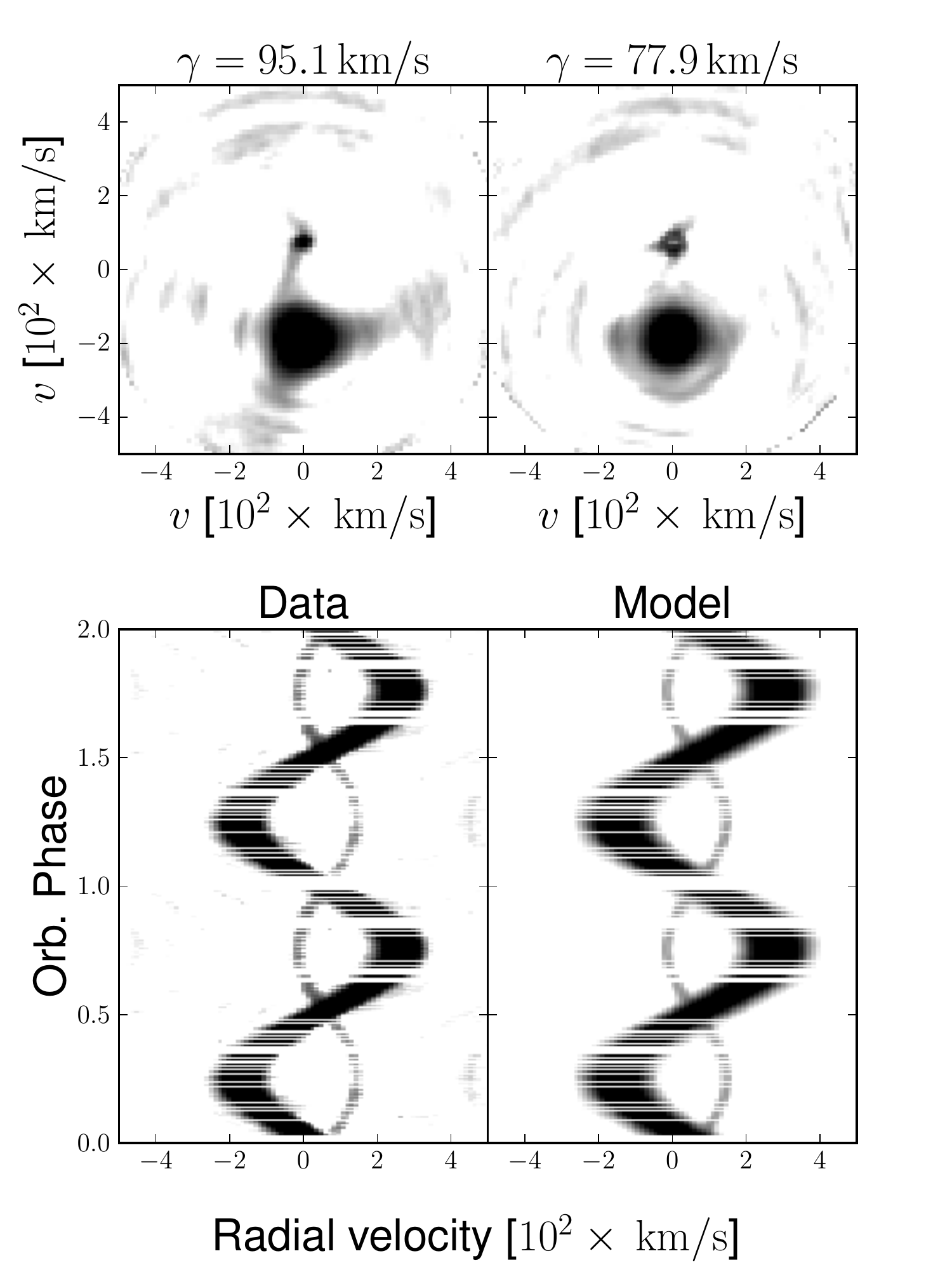}
 \caption{Doppler tomography of the \Ha~emission line. The Doppler tomograms are shown in the top left and right panels, obtained considering the solution for the $\gamma$-velocity of the extra component of the line (probably generated by the WD) and for the M-dwarf component, respectively. In the bottom panel we show a trailed spectrum of the data (left) and the resulting model based on the map with the $\gamma$-velocity of the additional component. See text for a more detailed explanation. }
\label{fig:dt_ha}
\end{figure}

In Fig.~\ref{fig:dt_ha} we show the resulting Doppler tomogram obtained with the maximum entropy reconstruction code provided by \cite{Spruit:1998p2514}. One of the fundamental aspects of Doppler tomography is that the lines must be first corrected for the $\gamma$-velocity of the system. For the \Ha~emission line of \object{RR~Cae}, we already know the existence of two components: one that traces the motion of the M dwarf and another that traces the WD. When fitting sinusoids to obtain the RV solution, we measured different $\gamma$-velocities for each component, the most reasonable explanation being that the WD component is redshifted because of the gravitational redshift of this star.

The overall effect of having components with different $\gamma$-velocities in a Doppler tomogram is that point-like features are replaced by rings with an asymmetric brightness distribution that depends in the distribution of observed phases. To study this effect on the Doppler tomogram of the \Ha~emission line, we obtained reconstructions with two different values of the $\gamma$-velocity. The results of the brightness distribution are labeled accordingly in Fig.~\ref{fig:dt_ha} (upper panels). The lower panels show a trailed spectra of the data (left) and the resulting model for the reconstruction with $\gamma = 95.1 {\rm km/s}$ (right), which provides a better view of the fainter WD component in the reconstruction. From the maps and trailed spectra it is clear that there are no other components to the \Ha~emission line, in addition to those of the M dwarf and the WD.

\subsection{Doppler imaging}
\label{sec:dopimg}

We applied indirect imaging techniques to map the surface brightness distribution of the M-dwarf component of the system. These techniques were successfully applied to several data sets to provide  surface images of rapidly rotating single and close binary stars. For the close and/or semi-detached binaries, tidal distortion of the imaged star by its companion requires an adaptation of the techniques -- also known as Roche tomography \citep{Watson:2001p1318}. 

The technique works by representing the tidally distorted surface of the star as a grid of uniformly distributed pixels, with a specific intensity assigned to each one. By changing the specific intensity of each pixel, we are able to map asymmetries in the rotationally broadened absorption profile which, in turn, are the result of temperature differences caused by the appearance of cool star spots. A more detailed description of the properties of our code can be found in \cite{2011AJ....142..106R}.

{
As most indirect-imaging techniques, Doppler imaging relies on a number of requirements that dictate the quality and reliability of the final surface brightness reconstruction. Two of the most important requirements are the signal-to-noise (S/N) of the data and knowledge of the system parameters: inclination of the system ($i$), masses of the two components ($\rm M_1$ and $\rm M_2$), radius of the imaged component (in this case $\rm R_2$) and the systemic velocity ($\gamma$). There are two important techniques designed specially to help improve the S/N and determine of system parameters, known as {\it least square deconvolution} (LSD) and {\it entropy landspace}, respectively. 
}

\subsubsection{Least square deconvolution}
\label{sssec:lsd}

When obtaining data for Doppler imaging one must be aware that there is a direct relationship between stellar rotation, spectral resolution, and exposure time to fulfill the technique's requirements for spatial resolution. This prevents, or at least discourages, long exposures to be used to obtain high S/N spectra. To circumvent this problem and allow the application of Doppler imaging to intermediate or low S/N spectra, \cite{Donati:1997p2174} proposed the LSD technique (see \citealt{Kochukhov:2010p3030} for a review). 

The LSD technique works by combining the multitude of absorption features in the spectrum into a single line with improved S/N. The improvement in S/N is scaled by the square-root of the number of lines weighted by their relative depth. In addition, the spectrum should be rectified by the continuum and a line list with central wavelengths and relative depths must be provided. For cool stars (${\rm T}\lesssim 4000\,{\rm K}$) one must also be cautious to avoid molecular bands in the stellar spectrum. 

\begin{figure*}
\centering
\includegraphics[width=0.8\textwidth]{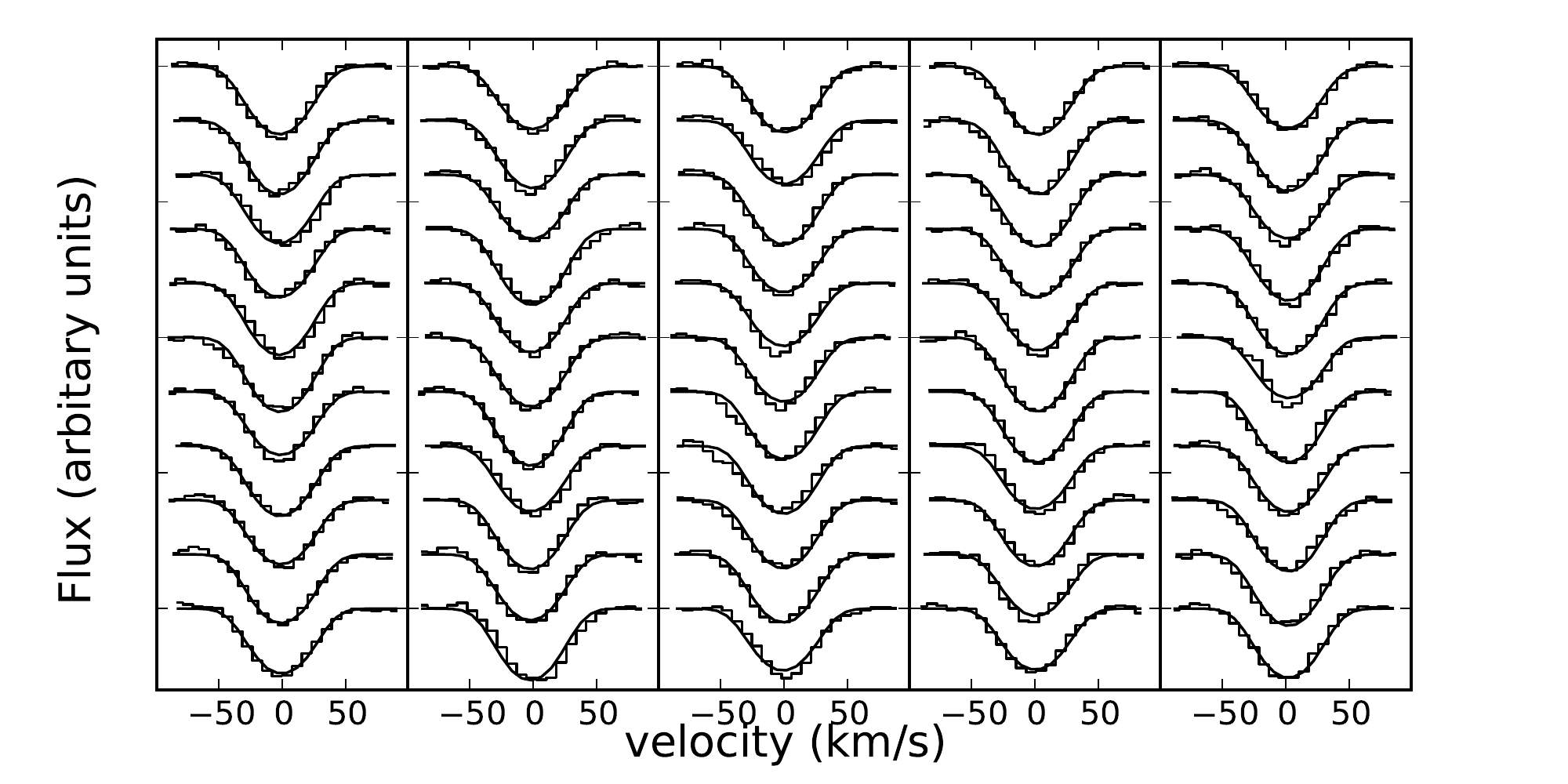}
 \caption{Selected LSD and maximum entropy profiles for \object{RR~Cae}. Only one for every two profiles are shown. The radial and $\gamma$-velocity of the M-dwarf star was removed from the profiles using the parameters listed in Table~\ref{tab:par}. The profiles are shifted in the y-direction for better visualization.}
\label{fig:lsdprof}
\end{figure*}

For \object{RR~Cae}, we selected four spectral regions in the mid and red spectra for applying the LSD technique. A summary of the spectral regions with the total number of lines and expected gain in S/N is shown in Table~\ref{tab:lsd}. A line list was retrieved from the VALD catalog \citep{Kupka:2000p3119, Piskunov:1995p139} for a stellar temperature and surface gravity of $\rm T = 3500\,K$  and $\log g (\rm cgs)= 5.0$, respectively, consistent with those of an M4-5V star \citep{Baraffe:1998p637}. { A collection of some of the resulting LSD profiles is shown in Fig.~\ref{fig:lsdprof} and a trailed spectrum with the resulting absorption profiles is shown in the left-hand panel of Fig.~\ref{fig:rochimg}.  Since the data must be continuum normalized, we did not use WD-subtracted spectra for obtaining the LSD profiles. Moreover, since only the redder part of the spectra was used in this procedure, the impact of the WD on the results is negligible.}

\begin{table}[h]
\caption{\label{tab:lsd} Summary of spectral regions used in the LSD and the expected gain in S/N.}
\centering
\begin{tabular}{ccc}
\hline\hline
Region [\AA]	& 	\# of lines	& 	S/N gain	\\
\hline
$4954:5165$	&	~49		&	~2.2		\\
$5240:5450$	&	~60		&	~2.3		\\
$5930:6145$	&	~75		&	~2.5		\\
$6165:6532$	&	138		&	~3.1		\\
Total		&	322		&	10.1		\\
\hline
\end{tabular}

\end{table}

\subsubsection{Entropy landscape}
\label{sssec:sland}

Another critical aspect of Doppler imaging is its dependency on precise system parameters. For binary stars as \object{RR~Cae}, the masses of the components in addition to the inclination, radius of the imaged star, rotation period, and systemic velocity ($\gamma$-velocity) must be known. Given the short timescale taken for tidal syncronization in compact binaries, we may consider that the rotation period and inclination of the M star are the same as those of the binary system. Since \object{RR~Cae} is a deep eclipsing binary, most of its system parameters are well defined. 

To improve previous estimate of system parameters and minimize the introduction of artifacts in the image reconstruction, we applied the entropy landscape technique \citep{1994A&A...288..773R}. This technique consists of performing the image reconstruction procedure for a set of system parameters, storing the final values of the entropy. The best system parameters are those that lead to the most featureless brightness distribution, i.e., the highest entropy reconstruction. The parameters that maximize the entropy of the reconstruction and were adopted throughout this analysis are listed in Table~\ref{tab:par}.

\subsubsection{Doppler imaging results}
\label{ssec:dires}

In addition to the system parameters listed above, our model also accounts for atmospheric effects such as gravity- and limb-darkening. We used the gravity-darkening coefficient $\beta = 0.08$, following the results of \cite{Lucy:1967p2997} for stars with deep convective envelopes. For the limb-darkening coefficient we adopted the nonlinear prescription of \cite{Claret:2000p1814},
\begin{equation}
\label{eq:limb}
\frac{I(\mu)}{I(1)} = 1 - a_1(1-\mu^{1/2}) - a_2(1-\mu) - a_3 (1-\mu^{3/2}) - a_4 (1 - \mu^2),
\end{equation}
where $I(\mu)$ is the specific intensity at $\mu$, the angle between the line of sign and the emerging flux\footnote{$I(1)$ is the specific intensity at the center of the stellar disk} and $a_n$ the $n-th$ limb-darkening coefficient (with $n$ going from 1 to 4, see Table~\ref{tab:par}). 

Determining the coefficients of the limb-darkening is not straightforward. In particular, the coefficients are calculated based on the stellar continuum, while here we are interested in the absorption lines. Nevertheless, the LSD profiles were constructed by combining a series of lines from a broad spectral region, weighting out the coefficients in a similar way as a broad-band filter. The spectral regions used to produce the LSD profiles approximately encompass the region defined by the B and V filters. Therefore, we used the unweighted mean of the B and V limb-darkening coefficients for ${\rm T_{eff} = 3500K}$, $\log g (\rm cgs) = 5.0$ and solar metallicity (the same as used for retrieving the line list in the LSD procedure). The resulting limb-darkening coefficients used for our analysis are listed in Table~\ref{tab:par}. 

\begin{figure*}
\centering
\includegraphics[width=0.45\textwidth]{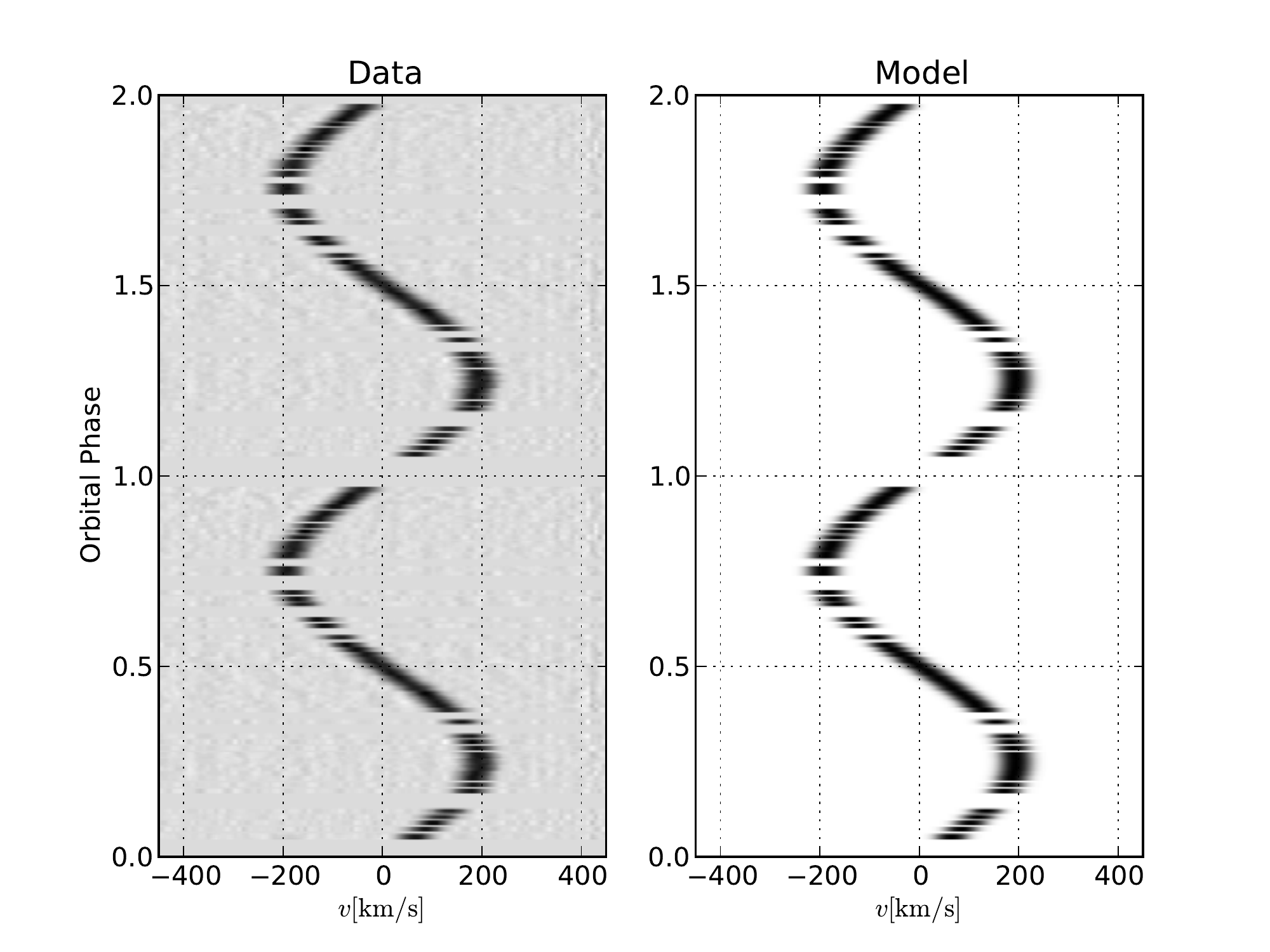}
\includegraphics[width=0.45\textwidth]{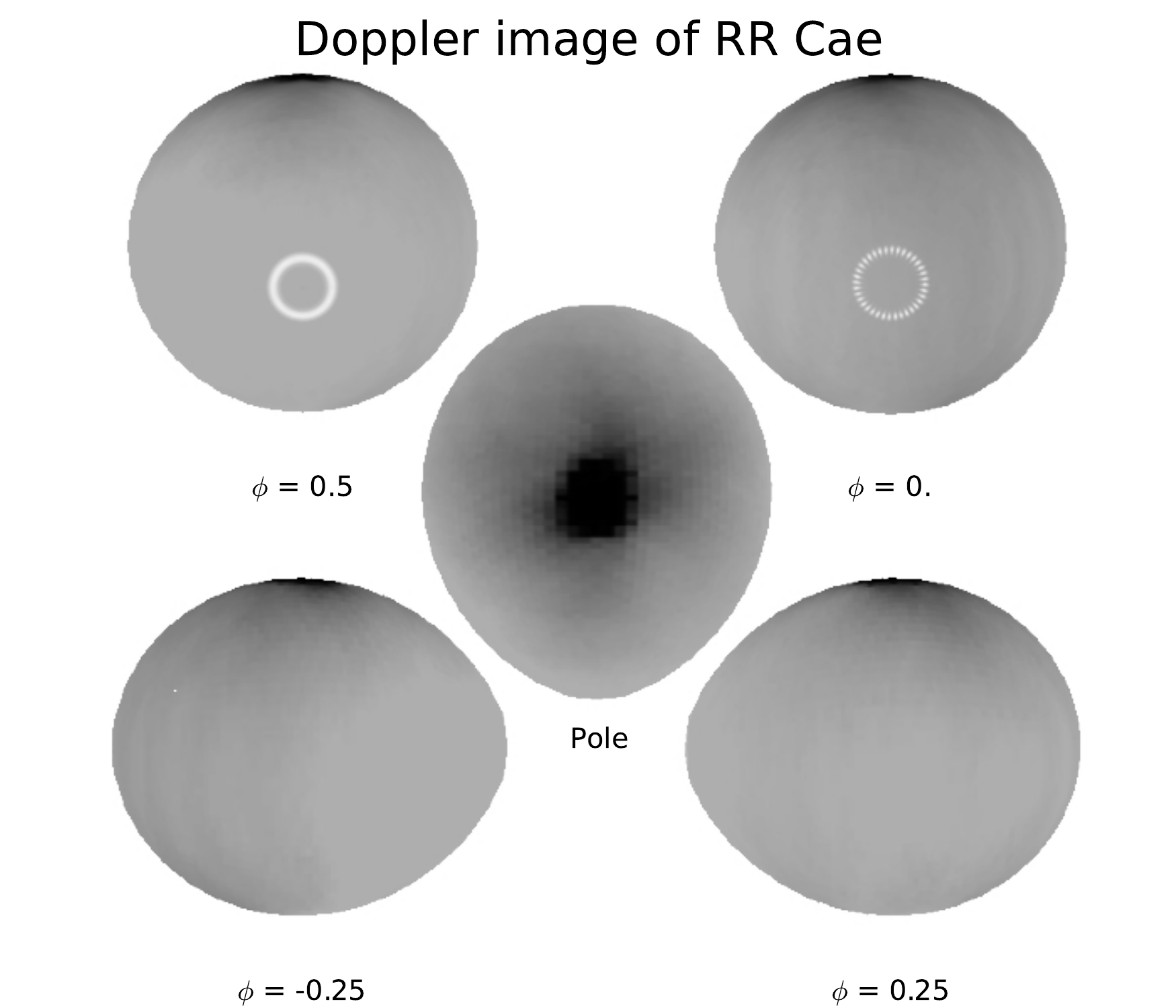}
 \caption{{\it Left:} LSD and modeled trailed spectrogram of the absorption features from the M-dwarf of \object{RR~Cae}. The profiles are repeated in phase for better visualization. {\it Right:} Doppler imaging reconstruction of the surface brightness distribution of the M-dwarf at four binary phases (corner maps) and as seen from the pole of the star (center map). The corresponding binary phase is indicated below each map. { A solid and dotted circle is overlaid around the point in the stellar surface that is crossed by the binary axis. The solid line represents the inner face of the star, i.e., the one facing the WD.} Here dark colors represents regions with lower specific intensity, and therefore, cooler or spotted regions.}
\label{fig:rochimg}
\end{figure*}

Surface brightness distribution of the M star of \object{RR~Cae} are shown in Fig.~\ref{fig:rochimg} at five different viewing angles. { The resulting maximum entropy line profile is overlaid on the LSD profiles in Fig.~\ref{fig:lsdprof} and is shown as a trailed spectra in the middle panel of Fig.~\ref{fig:rochimg}.} The orbital phase of each view is shown below each surface map. These maps are seen at the inclination angle determined by the entropy landscape technique except for the central image, which is viewed from the pole of the star. The most prominent feature in the surface brightness distribution is a polar spot. The spot smoothly extends to lower latitudes with a more extended feature in the trailing side of the star (best seen around phase 0 and -0.25, see Fig.~\ref{fig:rochimg}). 

Doppler imaging assumes that there is no variation in the surface brightness distribution of the imaged star with time. Since the spectra of \object{RR~Cae} were obtained at different epochs, we verified that the star was in a consistent brightness state. By analyzing the behavior of the \Ha~line, we showed (see sect.~\ref{ssec:act})  that there is a interchangeable small change in the activity state for 5 of the 16 epochs. For instance, the system appears to be in an increased magnetic activity state at EP01 and EP02, switching to a more basal and consistent activity level from EP03 to EP07. From EP08 to EP16 it seems that the star flickers between lower and higher states. At the same time, it is important to bear in mind that the \Ha~emission line likely originates from the upper layers of the stellar chromosphere \citep{Hall:2008p129}, and that these variations may well be caused by small magnetic reconnection events in these upper layers that will have only a mild effect on the actual starspot distribution at the stellar surface. 

To provide a careful analysis of the LSD profiles we started by considering only spectra taken between epochs EP03 to EP07, roughly 40 profiles. These data were obtained in five successive nights of observations between two epochs of enhanced activity (EP01, EP02 and EP08). Nevertheless, the data from EP03 to EP07 do not cover phases between [0.15:0.40] and [0.55:0.8], which prevented us from obtaining a reliable image reconstructions. We then complemented the initial data set with profiles from EP11 and EP15, which are epochs with ``normal'' activity levels. Indirect imaging of this data set leads to the surface brightness distribution shown in Fig.~\ref{fig:rochimg}.

Given the high inclination of the system ($i = 79\degr$), it is quite surprising that we were able to detect a polar feature in the surface brightness reconstruction. It is a well-known limitation of the technique that it cannot detect polar features at high stellar inclinations, since the polar regions are poorly sampled throughout the orbit in these cases. This is illustrated in Fig.~\ref{fig:rochimg}, which shows that the polar regions are seen from very high viewing angles at all phases.

This encouraged us to perform simulations to determine the ability of our reconstruction code to recover polar features at such high inclinations. For that we generated synthetic profiles for a set of binary inclinations with the same spectral and temporal resolution as our data and varying the S/N from $100$ to $500$ (the range we expect our final LSD profiles to be in). For the synthetic profiles we used different surface brightness distributions with and without polar spots. The results indicate that we are still able to detect polar features for inclinations as high as $i \lesssim 85\degr$, though with a significant loss in the brightness contrast of the respective feature at the higher inclination end. No polar spot artifact is introduced in the reconstructions for any of the cases where the original surface brightness distribution does not contain such a structure. Therefore, we are confident in the detection of the polar structure in the surface brightness of the M-dwarf component of \object{RR~Cae}.

\subsection{Accretion}
\label{ssec:acc}

One of the most intriguing aspects of the blue spectra of \object{RR~Cae} is the multitude of metal absorption features. Analysis of the kinematics of these lines over the orbital period suggests that they originate from the WD, and not from the MS component (the emission lines in the core of the Balmer lines, see Fig.~\ref{fig:sp2_H_avg}, do originate from the WD, however). The most likely interpretation for the heavy elements in the WD spectrum is accretion from the M-dwarf wind \citep{Debes:2006p3314}. A detailed abundance analysis of the WD photosphere can thus provide a good estimate of the wind-accretion rate (see below). 

To measure the surface abundance from the numerous absorption features, we first proceeded in redetermining consistently and independently the basic atmospheric parameters ($\rm T_{eff}$ and $\log g$) from our data. To obtain accurate parameters, we had to carefully attend to the possible contribution of the M dwarf in the blue part of the spectra. Using a method for subtracting the contamination from the secondary similar to that described in detail in \citet{2011ApJ...743..138G}, we fitted the Balmer lines from $\rm H\gamma$ to $\rm H8$ and obtained $\rm T_{eff}$ and $\log g$. In retrospect, since the contribution of the M-dwarf in the blue part of the spectrum was, at most, only a few percent, we found that the subtraction had no significant effect on the final solution (change of less than $\rm 100 K$ for $\rm T_{eff}$ Teff and 0.10dex for $\log g$). Our final results, $\rm T_{eff} = 7260 \pm 250K$ and $\log g = 7.8\pm0.10$, are slightly different from those of \citet[$\rm T_{eff} = 7540 K$ and $\log g = 7.7$]{Maxted:2007p4248}, but agree well within the uncertainties.

\begin{figure*}
\centering
\includegraphics[height=\textwidth,angle=270,trim = 2in 0in 0.45in 0in]{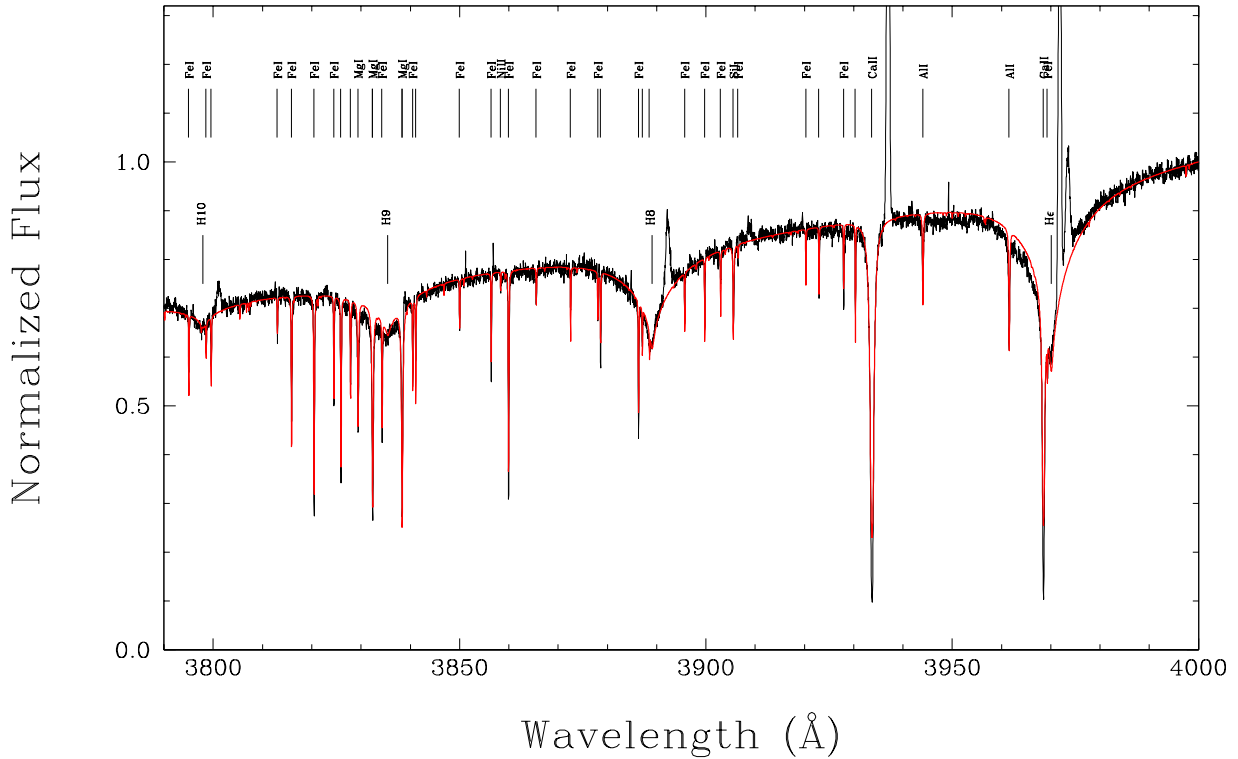}
\caption{An average of blue spectra of \object{RR~Cae} around phase $0.25$ corrected for the radial velocity of the WD (black) with the respective atmosphere model spectrum (red). }
\label{fig:sp_wdfit}
\end{figure*}

Next, we calculated a model structure using the atmospheric parameters determined above followed by grids of synthetic spectra with metal lines for each element of interest. Our fitting method is very similar to that described in detail in \citet{2012ApJ...749....6D}. In Fig.~\ref{fig:sp_wdfit} we show our best-fit model in the 3800-4000\AA~range while in Table~\ref{tab:wdfit} we present the abundances for all the detected elements in the WD photosphere. We find that the abundance of the different species, with respect to solar, are quite consistent with an average value of $<\log[X/X_\odot]> = -2.8 \pm 0.1$ (where the error is the standard deviation). It is interesting to note that by repeating this procedure for fitting metal lines but using instead the \citet{Maxted:2007p4248} $\rm T_{eff}/\log g$ solution, we still manage to reproduce the data quite well, but the dispersions in abundance for the individual fits of the numerous \ion{Fe}{} and \ion{Ni}{} lines are much higher ($\sim \rm 0.12\,dex$ versus $\rm \sim 0.05\,dex$ with our new solution). Furthermore, individual calcium abundance measurements from the \ion{Ca}{II} H\&K lines also agree better with that from \ion{Ca}{I}-4226\AA~line when using our new solution. We tested a few other $\rm T_{eff}/\log g$ couples near our preferred solution and found that the effective temperature and surface gravity that we determined from the Balmer lines allows the best overall fit for the many metal features. 

\begin{table*}
\caption{\label{tab:wdfit} Abundances of metals in the WD spectra of RR~Cae.}
\centering
\begin{tabular}{lccccc}
\hline\hline
Element	&\multicolumn{3}{c}{Abundance} &$\log(\tau_{D}[{\rm yr}])$&$\dot{\rm M}_{\rm acc}$ 	\\
	&$\log[X/H]$	&$\log[X_\odot/H]$&$\log[X/X_\odot]$& &$\rm M_\odot \cdot yr^{-1}$	\\
\hline
Mg	& $7.1\pm0.2$ & $-4.42$	&$-2.65$& $3.94$& $7.5 \times 10^{-16}$			\\
Ti 	& $9.8\pm0.3$ & $-6.98$	&$-2.77$& $3.85$& $7.0 \times 10^{-16}$      		\\
Ni 	& $8.5\pm0.3$ & $-5.75$	&$-2.79$& $3.79$& $7.6 \times 10^{-16}$      		\\
Fe 	& $7.2\pm0.3$ & $-4.50$	&$-2.70$& $3.76$& $1.0 \times 10^{-15}$			\\
Ca 	& $8.4\pm0.3$ & $-5.64$	&$-2.77$& $3.92$& $5.9 \times 10^{-16}$			\\
Al 	& $8.2\pm0.2$ & $-5.53$	&$-2.67$& $3.92$& $7.5 \times 10^{-16}$      		\\
Si 	& $7.2\pm0.2$ & $-4.45$	&$-2.76$& $3.95$& $5.7 \times 10^{-16}$      		\\
Cr 	& $9.4\pm0.2$ & $-6.33$	&$-3.11$& $3.83$& $3.3 \times 10^{-16}$      		\\
\hline
\end{tabular}

\end{table*}



Since the timescales for the diffusion of heavy elements in WDs are very short compared whith their evolutionary timescales, the detection of metals in its spectrum provides a direct evidence of accretion $\rm T_{eff}/\log g$. Because the M component of \object{RR~Cae} is still far from filling its Roche lobe, wind and/or debris accretion should provide observed metal enrichment at the WD photosphere. Following the analysis provided by \cite{2011A&A...532A.129T} of \object{LTT~560} {  and \citet{2012MNRAS.419..817P} of \object{SDSSJ1210}}, we investigated the wind-accretion rate required to match our model abundances. We started by assuming a steady accretion rate onto the WD which, from \cite{1993ApJS...84...73D}, provides a relation for the accretion rate 
\begin{equation}
\label{eq:wdacc}
\dot{\rm M}_{\rm acc} = \frac{{\rm M_{\rm wd} } q X}{\tau_D X_{\rm acc}}.
\end{equation}
Here $X$ is the abundance of a specie in the WD atmospheric layer, $X_{\rm acc}$ the abundance of the accreted material, $q$ is the mass fraction of the atmospheric layer and $\tau_D$ is the diffusion timescale. 

We used recently generated tables for convection zone models and diffusion timescales of metals in cool white dwarfs calculated by Fontaine et al. (in preparation). These are based on improved calculations of diffusion coefficients obtained by 1) using a much more accurate numerical scheme for estimating the so-called collision integrals, and 2) introducing a more physical prescription of the screening length used as the independent variable in evaluating these coefficients. Likewise, the physical model of pressure ionization used to compute the average charge of a diffusing element has been improved significantly and now leads to charges that vary smoothly with depth, as they should. We interpolated linearly in these tables to obtain the appropriate values for the WD parameters, in this case $\log q = -8.19$, while $\tau_D$ for each specie is shown in Table~\ref{tab:wdfit} together with the implied mass-accretion rate. The average value is $\rm \dot{M}_{acc} = 7\pm2 \times 10^{-16} M_\odot \cdot yr^{-1}$ (where the error is the standard deviation).

\cite{Debes:2006p3314} derived a somewhat lower value for the accretion rate of \object{RR~Cae} ($\rm \dot{M}_{acc} = 4 \times 10^{-16} M_\odot \cdot yr^{-1}$). Considering that this was based on \ion{Ca}{I} line alone and in a less consistent approach with respect to settling times and mass fraction of the atmospheric layer, our value probably provides a more precise estimate. 

A similar analysis performed by \cite{2011A&A...532A.129T} on LTT~560 {and on SDSSJ1210 by \citet{2012MNRAS.419..817P}} derived mass-accretion rates of $\rm \dot{M}_{acc} \sim 5 \times 10^{-15} M_\odot \cdot yr^{-1}$ for both systems. When comparing their results with that of \cite{Debes:2006p3314}, the lower orbital period of LTT~560 may well account for the lower accretion rate on \object{RR~Cae}, considering a Bondi-Hoyle accretion scenario where ${\rm \dot{M}_{acc} } \sim 1/a^2$ (where $a$ is the orbital separation). Accordingly, the wind accretion of LTT~560 { and SDSSJ1210} should be $\sim 3\times$ that of \object{RR~Cae}. Even though the mass-accretion rate still diverges by a factor of $2$, our higher estimate brings this interpretation in closer agreement, specially if we consider all uncertainties involved in its determination. 

\subsection{System parameters}
\label{ssec:spar}

{
Each subset of our data analysis provided estimates of different parameters of the system. It is interesting now to unify these determinations, check the internal consistency and compare them with those provided in the literature. In Table~\ref{tab:par} we list a compilation of the system parameters determined in this work and those of \citet{Maxted:2007p4248}, \citet{Bruch:1999p3205} and \citet{Bruch:1998p4261}.
}

\begin{table*}
\caption{\label{tab:par} System parameters provided by the entropy landscape technique, used in the Doppler image reconstruction, and a comparison with previously estimated values from the literature.}
\centering

\begin{tabular}{cccc}
\hline\hline
\multicolumn{4}{c}{Limb-darkening coefficients}				 \\
\hline
$a_1 = +0.406$	 & $a_2 = +1.004$	& $a_3 = -0.733$	& $a_4 = +0.192$	 \\
\hline
\end{tabular}

\vspace{0.25cm}

\begin{tabular}{ccccc}
\hline\hline
Parameter	 					& This work 	& \citet{Maxted:2007p4248} 	& \citet{Bruch:1999p3205}	& \citet{Bruch:1998p4261}				 \\
\hline
$i$		  				& $^\dagger79.0\degr$	& $76\degr-90\degr$ & $84.04\degr$ & $87.9\degr$    	\\
${\rm M_1}$				&  $^\dagger0.45 {\rm M}_\odot/^{\star}0.43\pm0.02 M_\odot$	& $0.440\pm0.023 \rm M_\odot$& $0.467\rm M_\odot$& 	$0.394\rm M_\odot$ \\
${\rm M_2}$				& $^\dagger0.15 {\rm M}_\odot$& $(0.182-0.183)\pm0.012\rm M_\odot$& $0.095\rm M_\odot$& 	$0.089\rm M_\odot$ \\
$\rm R_2$				& $^\dagger0.21 {\rm R}_\odot$& $(0.203-0.215)\pm0.015 \rm R_\odot$& $0.189\rm R_\odot$& $0.134\rm R_\odot$ 	 \\
$\gamma$				& $^\dagger80 \rm km \cdot s^{-1}/^\ddagger$& $^\P85.8 \pm 3.6 \rm km \cdot s^{-1}$ & $40\rm km \cdot s^{-1}$& $--$ 	 \\
$\log g_1$				& $7.8 \pm 0.1$& $7.97$& $--$& $--$ \\
$\rm T_1$					& $7260\pm250 \rm K$ & $7540\pm175 \rm K$ & $7005 \rm K$ & $7005 \rm K$\\
\hline
\end{tabular}

$\dagger$ - Entropy landscape.

$\star$ - Radial velocity fit provides gravitational redshift and WD spectra modeling provides $\log g$.

$\ddagger$ - See Table~\ref{tab:rv}.

$\P$ - For the M dwarf. See their Table 2.
\end{table*}

%


{
In Sect.~\ref{ssec:rv} we used the RV measurements of the different spectral features to determine the mass ratio $q=0.376\pm0.005$ of the system and the gravitational redshift of the WD. The masses of each component were also obtained as a result of the entropy landscape technique used for Doppler imaging, which maps into a mass ratio of $q = 0.34$. Unfortunately, the entropy landscape technique does not provide an estimate of the uncertainty in the measurement. Given that the entropy landscape is less sensitive to the mass of the primary star, an $8\%$ difference is acceptable. On the other hand, our value is considerably higher than that obtained by \citet{Bruch:1999p3205} and lower than that of \citet{Maxted:2007p4248}.

One of the main parameters of the Doppler imaging is the inclination of the system. This parameter has a strong impact on the radial and rotation velocity and, therefore, should be strongly constrained by the entropy landscape procedure. Our result ($i = 79\degr$) is consistent with the range stipulated by \citet{Maxted:2007p4248}, though they favored higher values ($i = 86\degr$, see their Fig.~7). The same is also true for the results of \citet{Bruch:1999p3205}, who obtained a best-fit value of $i=84.04\degr$ and \citet{Bruch:1998p4261} where $i = 87.9\degr$. We recall that all these values where contemplated during the entropy landscape procedure as well as the combination of masses for the two components. For an inclination of $84.0\degr$ we are still able to obtain results consistent with those presented in Sect.~\ref{sec:dopimg}, although with lower entropy. For inclinations higher than $85\degr$ it is not possible to achieve convergence in the imaging procedure. Therefore, it is very likely that the inclination of the system is lower than previously thought. 

{
The masses of the components were also independently determined with our data set, using results from the different analyses. With the entropy landscape technique we obtain $\rm M_1 = 0.45 M_\odot$ and $\rm M_2 = 0.15 M_\odot$. By fitting the RV of the spectral components of the system, we obtain a gravitational redshift of $v_{gr} = 20 \pm 1 \, \rm  km \cdot s^{-1}$. Combining this with $\log g = 7.8 \pm 0.1$, obtained by modeling the spectrum of the WD (Sect.~\ref{ssec:acc}), results in $\rm M_1 = 0.43\pm0.02 M_\odot$, consistent with the result obtained by the entropy landscape technique and with the WD mass obtained by \citet{Maxted:2007p4248}, as well as that of \citet{Bruch:1999p3205} and \citet{Bruch:1998p4261}.  
}

We were also able to measure the radius of the M dwarf with the entropy landscape technique, which gives $\rm R_2 = 0.21 R_\odot$.  Our result agrees well with that of \citet{Maxted:2007p4248} and \citet{Bruch:1999p3205}, but is considerably higher than that of \citet{Bruch:1998p4261}. According to the \citet{Baraffe:1998p637} models, the M dwarf in \object{RR~Cae} would be up to $\sim 20\%$ oversized for its mass. Even though this is a considerable factor, it is not as high as those observed in partially convective stars, which may reach up to twice the expected radius \citep{2011AJ....142..106R,Vida:2009p758}.

}

\section{Discussion}
\label{sec:disc}

Doppler imaging of the M dwarf of \object{RR~Cae} revealed a high-latitude polar spot. The current stellar dynamo theory suggests that star spots are generated by the emergence of magnetic tubes when strong toroidal fields at the bottom of the convection region reach the buoyancy limit. Spots are supposed to emerge at equatorial regions of the star and then be dragged by strong Coriolis force, due to fast rotation, to the pole of the star \citep{Granzer:2002p1446}. Similar features have already been reported for detached DA + M-K dwarf \citep{Hussain:2006p696,2011A&A...532A.129T}, semi-detached systems \citep{Watson:2006p1413,Watson:2007p409}, and a variety of RS~CVn binaries (see \citealt{2009A&ARv..17..251S} for a collection of results of Doppler imaging of stellar surfaces). However, only a minority of these studies were performed for systems with fully convective components. 

Indeed, before this work, LTT~560 \citep{2011A&A...532A.129T} was the only fully convective M dwarf with a DA companion imaged with this technique. The combined result for \object{RR~Cae} and LTT~560 suggest that fast rotation plus fully convective stellar structures are equally capable to produce polar features such as those observed in partially convective stars, which are characteristics of large-scale magnetic fields. Furthermore, \citet{Morin:2010p3133} presented a spectropolarimetric survey of fully convective M dwarfs with magnetic field topologies obtained through Zeeman-Doppler imaging. Our results agree well with their detection of a strong large-scale toroidal axisymmetric magnetic field in an M8 dwarf (V10), thus indicating that this type of magnetic field configuration may indeed be common at the lower-mass end of the fully convective regime ($\rm M < 0.2 M_\odot$).

It is interesting to analyze the activity index indicators in view of the {\it plage-} and {\it prominence}-like features proposed by \cite{1992AJ....104.1942H,1994AJ....107.1149H}. In short, these authors suggested that plages are ground-level chromospheric structures while prominences are vertically extended features. Even though both features are equally connected with chromospheric activity, the separation between them can provide hints on the overall geometry/configuration of the stars magnetic field. In Sect.~\ref{ssec:act} we calculated the flux ratio $\rm E_{H\alpha} / E_{H\beta} \sim 1.2$ for the M dwarf of \object{RR~Cae}. This low ratio ($\rm E_{H\alpha} / E_{H\beta} < 3$) is characteristic of more plage-like features, as noted by \cite{1992AJ....104.1942H} and also observed in the Sun \citep{1991PhDT.........9C}. This result is also supported by the Doppler tomogram of the \Ha~line presented in Fig.~\ref{fig:dt_ha}, where we observe no extended features.

These results are consistent with those for the similar system LTT 560 \cite{2011A&A...532A.129T}, but contrast with observations of QS Vir by \cite{Ribeiro:2010p2079} and \cite{2011MNRAS.412.2563P}. In addition to the orbital period, the main difference between \object{RR~Cae} and LTT~560 or QS~Vir is that the M dwarf in the latter has a larger mass and, presumably, earlier spectral type. At the same time, prominence-like features are commonly observed in RS~CVn binaries (evidence is abundant in the literature but see, for instance, \citealt{1992AJ....104.1942H,1994AJ....107.1149H} and references therein), though mostly on early-K to late-F dwarfs and giant components. \cite{2011A&A...531A.113T} also observed a number of additional components to the \Ha~emission line in detached DA+M binaries. The lack of more time-resolved and wavelength coverage\footnote{Their data reach from 5750-7230 \AA, thus missing \Hb~(4863\AA), which prevents them from using the flux ratio $\rm E_{H\alpha} / E_{H\beta}$ to identify prominence-like features.} resolution prevents us from distinguishing between prominence-like and WD-originated features.

Prominence-like features were also revealed in interacting magnetic CVs containing a M-dwarf star as late as (or even later than) those in \object{RR~Cae} and LTT~560 e.g. \cite{Kafka:2010p3121,2008ApJ...688.1302K}. Nevertheless, in these cases it is possible that the interaction of the strong magnetic field of the WD provides the mechanism for long-lasting stable prominences. Our result is also in line with recent investigation of coronal structures by \citet{2012MNRAS.424.1077L}. These authors demonstrated that in these lower-mass stellar regimes the corona is dominated by more compact structures originating close to the stellar surface. Furthermore, we are unaware of any evidence of long-lasting prominence-like features on M dwarfs with spectral types as late as those of \object{RR~Cae} and LTT~560.  

The large-scale prominence-like features may be responsible for the strong magnetic breaking observed on partially convective stars. If this interpretation is correct and fully convective stars are dominated by strong poloidal fields with small-scale plage-like structures, we would be able to explain the strong difference in breaking scales for stars in the different regimes \citep{2008AJ....135..785W}. Most importantly, it would also provide strong evidence in favor of the disrupted-braking model for the evolution of CVs \citep[and references therein]{2008arXiv0809.1800R}. In this case, the magnetic field of the donor star switches to a more plage-dominated state with less efficient magnetic-braking properties. 

Another important parameter that affects the breaking timescales is the wind mass-accretion rate. It may be that for some yet unknown reason, low-mass stellar winds are just not dense enough to provide efficient angular momentum losses. In this regard, \cite{Debes:2006p3314} estimated the wind-accretion rate of M dwarfs by analyzing the metallicity of \ion{Ca}{} in WD companions; \object{RR~Cae} was part of their sample. We improved the \cite{Debes:2006p3314} estimate of the \object{RR~Cae} wind-accretion rate with a more detailed analysis, which includes many more chemical species and employs a consistent method for calculating the WD metallicity, mass fraction of the atmospheric layer, and diffusion timescale. A similar analysis was performed by \cite{2011A&A...532A.129T} for LTT~560, revealing a higher, but still consistent, mass accretion rate. Altogether, there is no evidence that supports a drastic decrease in the wind-accretion rate for low-mass stars. 

RR Cae and LTT 560 show a remarkable similarity of stellar component parameters, e.g., the temperature and masses of their WD and M dwarf. This suggests that they share a similar age. The WD cooling age of the system is $7 \times 10^8 \, \rm years$, using the cooling tracks provided by \citet{2011ApJ...737...28B},\citet{2011ApJ...730..128T}, \citet{2006ApJ...651L.137K} and \citet{2006AJ....132.1221H}. 

On the other hand, the fact that LTT~560 has a much shorter orbital period suggests differences in their progenitor binaries. Recent simulations of post-common-envelope binary evolution by \citet{2012MNRAS.419..287D} showed that the main  parameters of the progenitors are indeed conserved in the binary evolution, thus suggesting that the stellar components of the two systems very likely shared similar properties (e.g., masses), but that LTT~560  originally had a shorter orbital period.

\section{Summary and conclusions}
\label{sec:end}

We provided a detailed analysis of the magnetic activity and accretion in the DA+M binary \object{RR~Cae}. Our results suggest a polar feature in the brightness distribution of the fully convective M-dwarf component, similar to that observed in other fast-rotating stars. At the same time, we found evidence that the surface of the active star of this system is dominated by low scale plage-like features in contrast to large-scale magnetic prominences. 

State-of-the-art atmospheric modeling algorithms were used to measure the properties of the WD component, which are $\rm T_{eff} = 7260\pm250K$, $\log g =7.8 \pm 0.1 {\rm dex}$ and,  $<\log[X/X_\odot]> = -2.8 \pm 0.1 {\rm dex}$. By considering the diffusion scenario, we were also able to estimate the mass-accretion rate required to sustain the observed metallicity of the WD, $\rm \dot{M}_{acc} = 7\pm2 \times 10^{-16} M_\odot \cdot yr^{-1}$.

\acknowledgement 

We would like to thank the anonymous referee for carefully revising our manuscript. TR acknowledge financial support from Conselho Nacional de Desenvolvimento Cient\'ifico e Tecnol\'ogico (CNPq/Brasil), Laborat\'orio Nacional de Astrof\'isica (LNA) and Instituto Nacional de Ci\^encia e Tecnologia de Astrof\'isica (INCT-A). This work is based on data products from observations made with ESO Telescopes at the La Silla Paranal Observatory under program ID 076.D-0142.

\bibliographystyle{aa}
\bibliography{../bibtex/papers} 

\end{document}